\documentclass[prb,aps,preprint]{revtex4}
\usepackage{graphicx}
\usepackage{dcolumn}
\usepackage{bm}
\usepackage[utf8]{inputenc}

\begin{document}
\title {Trap-assisted space charge limited transport in short channel MoS$_2$ transistor}
\author{Subhamoy Ghatak,$^1$}
\email{subhamoyghatak@gmail.com}
\author{Arindam Ghosh$^1$}
\vspace{1.5cm}
\address{$^1$ Department of Physics, Indian Institute of Science, Bangalore 560 012, India}


\begin{abstract}
We present temperature dependent $I-V$ measurements of short channel MoS$_2$ field effect devices at high source-drain bias. We find that
although the $I-V$ characteristics are ohmic at low bias, the conduction becomes space charge limited at high $V_{DS}$ and existence of an
exponential distribution of trap states was observed. The temperature independent critical drain-source voltage~($V_c$) was also determined. The
density of trap states was quantitatively calculated from $V_c$. The possible origin of exponential trap distribution in these devices is also
discussed.
\end{abstract}


\maketitle

The MoS$_2$-based transistors have drawn a considerable interest in recent years because of their wide variety of applications like high on/off
ratio transistor~\cite{single}, phototransistor~\cite{Yin}, non-volatile memory device~\cite{nonvolatile}, tunnel barrier
transistor~\cite{grws2gr,grmos2gr}, photodetector~\cite{photodetector} \emph{etc}. Although the device architectures with MoS$_2$ has been
well-explored, there are only few studies on the fundamental aspects of these devices like low mobility, localized electronic states, strongly
disordered electronic landscape. The localized nature of electronic states has been reported by different groups and was explained either by
variable range hopping~(VRH)~\cite{natureofelectronic,jariwala} or temperature dependent activated behavior~\cite{nanopatch,metalinsulatormos2}
but the origin of such strong disorder remained unaddressed. It was suggested that the substrate trap charges might be responsible for the
localization~\cite{natureofelectronic} because ultrathin flakes are often found to be influenced more by substrate induced external potentials.
On the other hand many similar low mobility systems such as organic semiconductors~\cite{polyvenyl,Berleb} and reduced graphene oxide
transistors~\cite{khandekerrgo} have been investigated where traps were found to originate from bulk due to structural inhomogeneity inside the
materials. The present work was motivated to find whether the manifestation of strong localization in MoS$_2$ is substrate effect or there is
some generic source of disorder in the bulk of MoS$_2$.

It is well-known that distribution of trap states can be addressed by measuring $I-V$ characteristics at high source-drain bias when device
characteristics are determined by space charge limited conductivity~(SCLC). The SCLC occurs when the injected carrier density($n$) at high
$V_{DS}$ exceeds the intrinsic carrier density($n_0$) of the material at that temperature~($T$). For a trap free solid, the SCLC theory predicts
that although the $I-V$ characteristic is ohmic at low $V_{DS}$, it changes to $J_{MG}=\frac{9}{8} \mu \varepsilon_0 \varepsilon_r
\frac{V^2}{L^3}$, which is known as Mott-Gurney relation~($J$ is current density whereas $\mu$, $\varepsilon_r$ and $L$ are mobility, relative
permitivity and length of the solid between two electrodes). In the presence of exponential trap charge distribution $h(E)=\frac{N_t}{k_B T_c}
e^{-E/k_B T_c}$ in the solid, the current-voltage relation becomes~\cite{mark},
\begin{equation}
J_T=\frac{\mu N_c}{q^{m-2}}  \left(\frac{2m-1}{m}\right)^{m} \left(\frac{\varepsilon_0 \varepsilon_r (m-1)}{N_t m}\right)^l \frac{V^m}{L^{2m-1}}
\end{equation}
where $N_t$ and $T_c$ are the total trap density and characteristic temperature of the trap distribution, $N_c$ is the effective density of
states in conduction band, $m=\frac{T_c}{T}+1$. When the measurements are done at $T\leq T_c$, $m\geq$2. Therefore the exponent becomes 2 at
$T_c$ and monotonically increases as $T$ decreases below $T_c$. This technique has already been widely used to explore the nature of trap states
in past particularly for low mobility devices~\cite{khandekerrgo,mottgurneykumer,mottgurneymorpugo}. In this study, we perform output
characteristics~($I-V$) of short channel MoS$_2$ transistors at high source drain bias i.e. longitudinal electric field. We find that at low
$V_{DS}$ the $I-V$ characteristics are ohmic i.e. $I_{DS}\propto V_{DS}^m$, where $m=1$ but as the electric field increases $m$ increases from
one and reaches $m\approx 2$ at room temperature. The exponent $m$ further increases monotonically as temperature decrease, indicating a trap
dominated SCLC.

Single layer of MoS$_2$ sheets were exfoliated using bilk MoS$_2$ crystals~(SPI supplies) on 285 nm degenerately doped Si/SiO$_2$ wafers by
micromechanical cleavage technique using scotch tape. Before exfoliation, the Si/SiO$_2$ wafers were thoroughly cleaned by RCA1~(5:1:1 of
H$_2$O:NH$_4$OH:H$_2$O$_2$) solution for 15 minutes and then in acetone and isopropyl alcohol in a ultrasonic bath. The layer number were
identified using optical microscope color contrast~\cite{visibilitygomez} and Raman spectroscopy~\cite{Ramanacsnano}. The devices were prepared
by ebeam lithography, metallization of 50~nm Au and liftoff in hot acetone. The channel lengths were kept smaller(typically from 80-200~nm) to
achieve high longitudinal electric field at relatively lower $V_{DS}$. A schematic of the 2-probe device is shown in Figure~1a whereas the inset
of Figure~1b shows the SEM image of Dev1~(see table). The transfer characteristic~($I-V_{BG}$, shown in Figure~1b, was measured using lockin
technique. From the transfer characteristics, all the devices were found to be highly $n$-doped even at $V_{BG}=0~V$~ and the threshold
voltage~($V_{T}$) was below $V_{BG}~=~-40~V$ in most of the devices~(see supplementary Figure~S1 and Table~1). The field effect
mobility~($\mu_{FE}$) was calculated from the slope of $I-V_{BG}$ graph~(Figure 1b) for all the devices which came out to be about 0.3-2
cm$^2$/Vs~(see Table~1).


The output characteristics~($I-V$) were measured at various temperature using two DC voltage sources~(Keithley 2400) acting as $V_{DS}$ and
$V_{BG}$. Figure~1c shows the $I-V$ characteristics near room temperature at different $V_{BG}$. It is clear from the graph that the $I-V$
characteristics are highly symmetric and linear for $V_{DS} \leq 100$~mV~(see supplementary for detail). This can be attributed to the
quasi-ohmic nature of the contacts between MoS$_2$ and Au, already discussed in literature~\cite{channellengthscaling}. It has also been
recently reported from a density functional theory~(DFT) calculation that the interface of MoS$_2$-Au contact becomes metallic which can give
rise of ohmic contact~\cite{TiorAupopov}. Moreover, experimental determination of work function reveals that for single layer MoS$_2$, its work
function reaches 5.1~eV from the bulk value ~4.5~eV, which is similar to the work function of Au~\cite{lightmatter}, indicating a high
possibility of ohmic contact at the interface. In our experiment, we see that the $I-V$ characteristics slowly become non-linear as $V_{DS}$
increases but always remain symmetric indicating absence of dominant Schottky barrier at the junction. Although, many experimental studies have
recently reported current saturation at high $V_{DS}$ in presence of high gate electric field~\cite{integratedbilayer,Qui,konar}, we saw a
strong increase in current after the ohmic regime till breakdown~(see supplementary). Figure~1d represents output characteristics till
$V_{DS}~=~2$~V at $V_{BG}$~=~0~V at different $T$. Two important things to note are as follows:~(i)~the $I-V$ characteristics are symmetric with
a non-linearity which increases as $T$ decreases and (ii)~strongly $T$ dependent in the experimental temperature range. These observations
effectively eliminate any significant effect of Schottky barrier or field induced tunneling~(direct or Fowler-Nordheim tunneling) in the
conduction process.


The room temperature $I-V$ characteristics of three different devices are plotted in Figure~2a in log-log scale. We found that for $V_{DS}\leq
500$~mV, $I_{DS} \propto V_{DS}^m$ with $m=1$, indicating ohmic conduction in the sample. With further increase of $V_{DS}$ the curves deviated
from linearity and became $I_{DS} \propto V_{DS}^m$ with $m\approx 2$. This suggest an initiation of trap-free space charge limited conductivity
regime. Although space charge limited conductivity may originate from drain induced barrier lowering~(DIBL) in short channel devices, we exclude
their dominance owing to the recent report~\cite{channellengthscaling}, showing extreme robustness of MoS$_2$ devices against short channel
effects. The temperature dependent of $I-V$ characteristics for Dev 5 are shown in Figure~2b. We found the slope of the curve at $T~=~285$~K to
be 1.7, which is little less than theoretically predicted value $m$~=~2. We also saw the exponent $m$ to be little lower than 2 in few other
devices near room temperature. Probably this happens because room temperature is greater that $T_c$ for these devices i.e. the devices have
higher free electron concentration than the trapped ones. We see that as temperature goes down to 205~K, the exponent $m$ reaches 2 and
monotonically decreases with further lowering of temperature and reaches 3 at 105~K~(Figure~2c). These observations clearly reveals existence of
exponentially distributed trap states in the ultra-thin MoS$_2$ devices.

\begin{table*}
\caption{\label{table1}Details of the devices:}
\begin{ruledtabular}
\begin{tabular}{cccccc}
 Device & Dimension (L$\times$W) \footnote{both dimensions in $\mu$m}& $V_{T}$ \footnote{in V~(approx.)}&$\mu_{FE}$\footnote{in cm$^2$/Vs at 295~K~(approx.)}& $V_c$~(V) & $N_t$~(10$^{17}$cm$^{-3}$) \\  \hline
 Dev 1 & $0.08 \times 1.6$ &\textgreater \textgreater~-~40&0.7&  3 & $1.6$  \\
 Dev 2 & $0.09 \times 1.7$ &\textgreater \textgreater~-~40&1.4& 3.2 & $1.7$ \\
 Dev 3 & $0.17 \times 1.9$ &$\sim$~-~15&0.4&  - & - \\
 Dev 4 & $0.16 \times 1.9$ & \textgreater~-~40 &0.5& 3 & $1.6$ \\
 Dev 5 & $0.22 \times 2.0  $ & $\sim$~-~40 &0.8& 4.3 & $2.3$\\
 Dev 6 & $0.16 \times 1.3 $ & $\sim$~-~70 &0.7&2.8 & $1.5$\\
\end{tabular}
\end{ruledtabular}
\end{table*}

One of the key aspect of trap limited space charge conductivity is that the density of the trap states~($N_t$) can be quantitatively determined
from the experimentally measured $I-V$ characteristics. It is believed that as $V_{DS}$ increases the trap states slowly get filled up by
injected charge carriers. At a critical $V_{DS}$, conductivity of the sample become independent of $T$ i.e the $I-V$ characteristics of
different temperatures intersect each other. This critical voltage is called $V_c$ and given by~\cite{mottgurneykumer},
\begin{equation}
V_c=\frac{qN_tL^2}{2\varepsilon_0 \varepsilon_r}.
\end{equation}

Therefor $V_c$ can be obtained by extrapolating the $I-V$ curves at different $T$. This has been illustrated in Figure~2c, where we extrapolate
the $I-V$ characteristics at $T$=205, 163, 124, 105~K. The extrapolated lines intersect each other at a critical $V_{DS}$=~4.3~V and trap
density can be estimated to be $\sim 2.3 \times 10^{17}$ cm$^{-3}$~for Dev 5~(see the table for other devices). We believe that these traps
originate from the bulk of the crystal and they are not interface trap states as discussed in a previous study~\cite{natureofelectronic}.

We additionally study how the $I-V$ characteristics change with increasing free carrier density in the channel using the backgate. It is obvious
that as the free carrier density increases more and more trap states will be filled at $V_{DS}$=0~V. As a result the effective density of trap
states will reduce. Therefor $V_c$ will decrease with increasing $V_{BG}$ owing to the equation~2. This has been illustrated in Figure~3, where
we plot the $I-V$ characteristics of the same device at $V_{BG}$~=~0, 20 and 40~V. The critical source-drain voltage $V_c$ was calculated for
$V_{BG}$~=~0 and 20~V by extrapolating the $I-V$ curves at 80, 110, 140, 175K yielding $V_c$=~2.8 and 2.2~V respectively~(Figure 3a,b). With
further increase of $V_{BG}$ to 40~V, the $I-V$ curves intersect each other at 1.4~V(Figure 3c). We did not include the 230, 262~K data because
we believe that at these elevated $T$~(typically $T~>~$200~K) the free carrier density slowly exceeds the trapped ones and calculation of $V_c$
from equation~1 is not appropriate anymore. This can be easily understood from Figure~3c, where $I-V$ characteristics at $T$~=~80, 110, 140 and
175~K intersect each other, the graphs at 230 and 262~K show no signature of intersection till $V_{DS}$=~2~V.


The origin of trap states in ultra-thin MoS$_2$ transistor remain controversial till date. Experimentally it was found that the charge transport
is governed by charged impurity scattering~\cite{natureofelectronic} and presence of a high-$k$ dielectric environment can improve the mobility
by orders of magnitude~\cite{single,dualgatemos2ieee}, which leads to a prediction that the uncompensated dangling bonds at SiO$_2$ surface acts
as trap states and strongly affects the carrier conduction in MoS$_2$~\cite{natureofelectronic}. However, our present measurement confirms that
apart from the substrate there is an exponentially distributed trap states in MoS$_2$ which might be more influential for carrier localization.
To explore this quantitatively we use the value of $N_t$ and calculate the correlation energy~($T_0$) of variable range hopping, known to be the
transport mechanism in ultrathin MoS$_2$ devices~\cite{natureofelectronic}. Assuming that the trap charges are distributed in an energy
bandwidth of $eV_c$, we find $T_0~=~\frac{1}{k_B \xi^3 (N_t/eV_c)}$ to be $2\times 10^5$~-~$2\times 10^4$~K using typical localization length
$\xi \approx 10-20$~nm, which agrees well with experimentally observed $T_0$ values~\cite{natureofelectronic}. It is believed that exponential
trap distribution arises due to surface defect or structural disorder in the bulk of the sample~\cite{surfacescience,khandekerrgo}. Similarly
inhomogeneity in MoS$_2$ may arise from external doping entity or point defects such as sulphur
vacancies~\cite{superconductingdome}~(responsible for $n$-type doping) in crystal lattice. Moreover, it has also been predicted that presence of
atmospheric oxygen can lead to a minute oxidation of MoS$_2$ to form MoO$_3$ at the surface~\cite{MoO3raman} which can lead to structural
disorder giving rise to trap states as found in case of reduced graphene oxide transistors~\cite{khandekerrgo}.

In conclusion, we study temperature dependent $I-V$ characteristics in short channel MoS$_2$ transistors at high $V_{DS}$ i.e. longitudinal
electric field. We find that the conduction is space charge limited at high $V_{DS}$ with an exponentially distributed trap states. The trap
density was qualitatively calculated measuring the temperature independent critical voltage $V_c$. We also discuss possible origin of such
exponetial trap distribution.

\begin{center}
\begin{figure}

\includegraphics[width=0.8\linewidth]{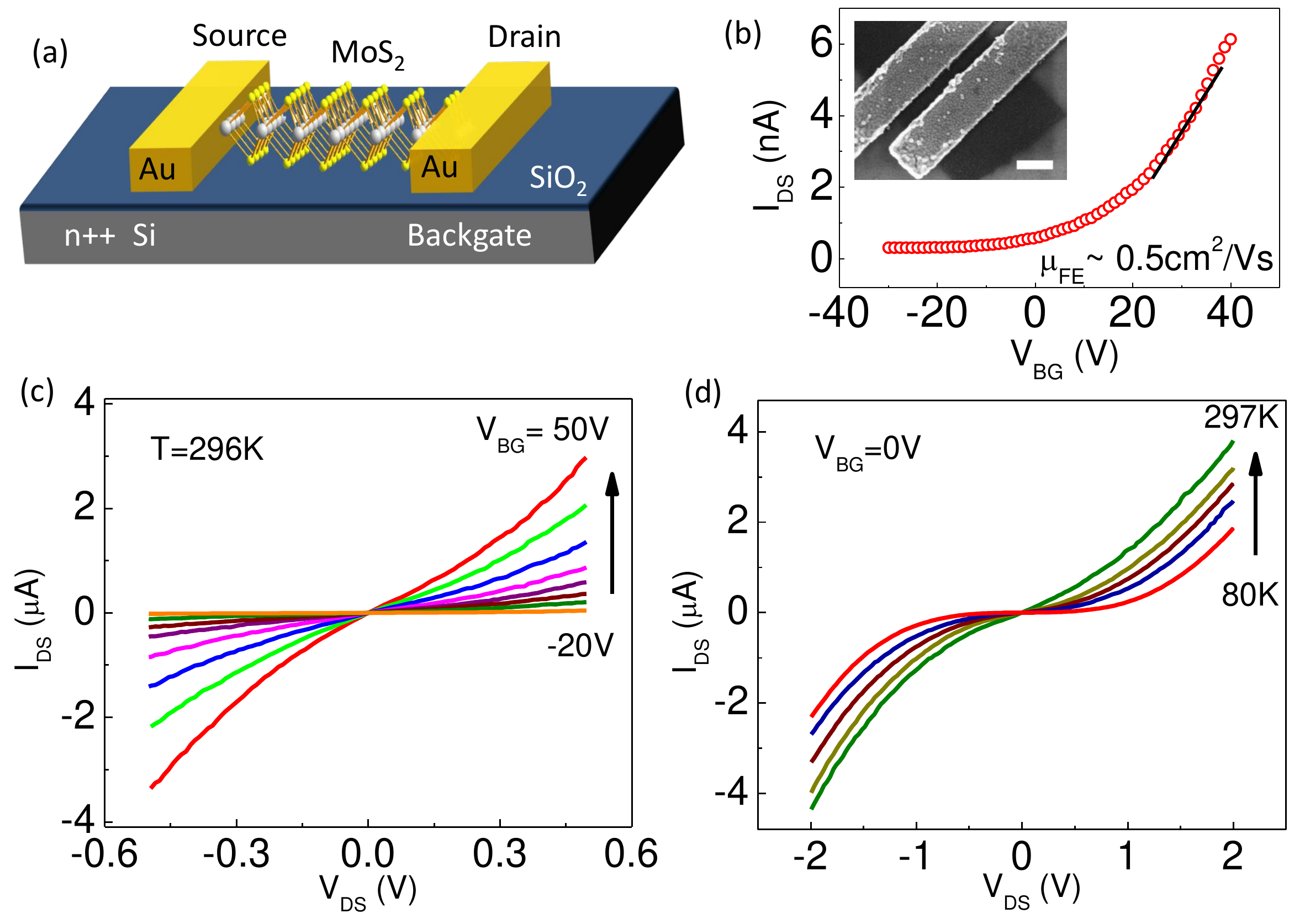}
\caption{(a)~Schematic of a 2-probe single layer MoS$_2$ device on Si/SiO$_2$ wafer. (b)~Transfer characteristic of Dev~2 at room temperature
and $V_{DS}=10$~mV. Field effect mobility was calculated from the slope of the curve as shown with the black line. Inset shows the SEM image of
Dev~1 with scale bar 150~nm. (c)~$I-V$ characteristics of Dev~6 at room temperature and $V_{BG}$ from 50~V to -20~V with 10~V gap. (d)~$I-V$
characteristics of Dev~4 at $V_{BG}$=0~V at 80, 160, 200, 240 and 297~K.}

\end{figure}
\end{center}
\newpage
\begin{center}
\begin{figure}

\includegraphics[width=0.5\linewidth]{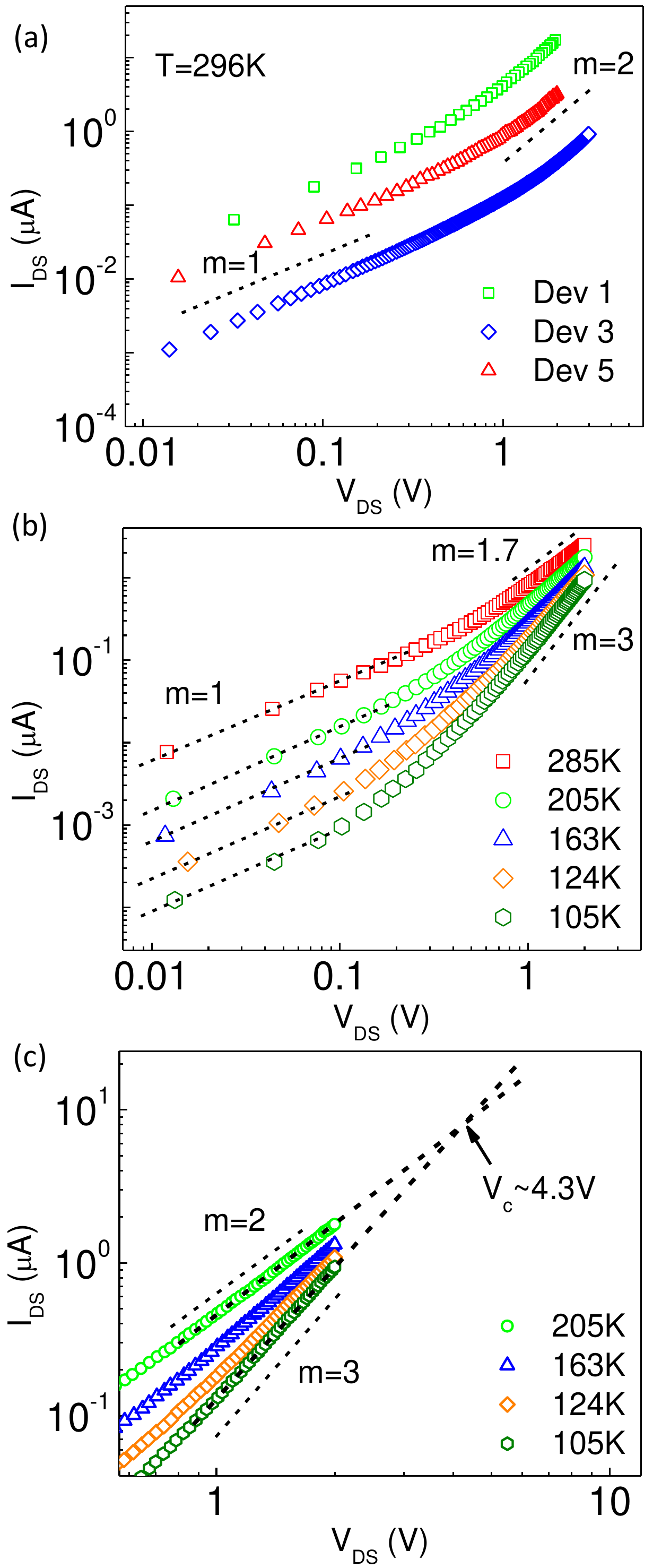}
\caption{(a)~$I-V$ characteristics of three different devices at room temperature and $V_{BG}$~=~0~V. (b)~$I-V$ characteristics of Dev~5 at
different temperatures at $V_{BG}$~=~10~V. (c)~Extrapolation of 105, 124, 163, 205~K data derived from Figure 2b to extract $V_c$.}

\end{figure}
\end{center}

\newpage
\begin{center}
\begin{figure}
\includegraphics[width=1\linewidth]{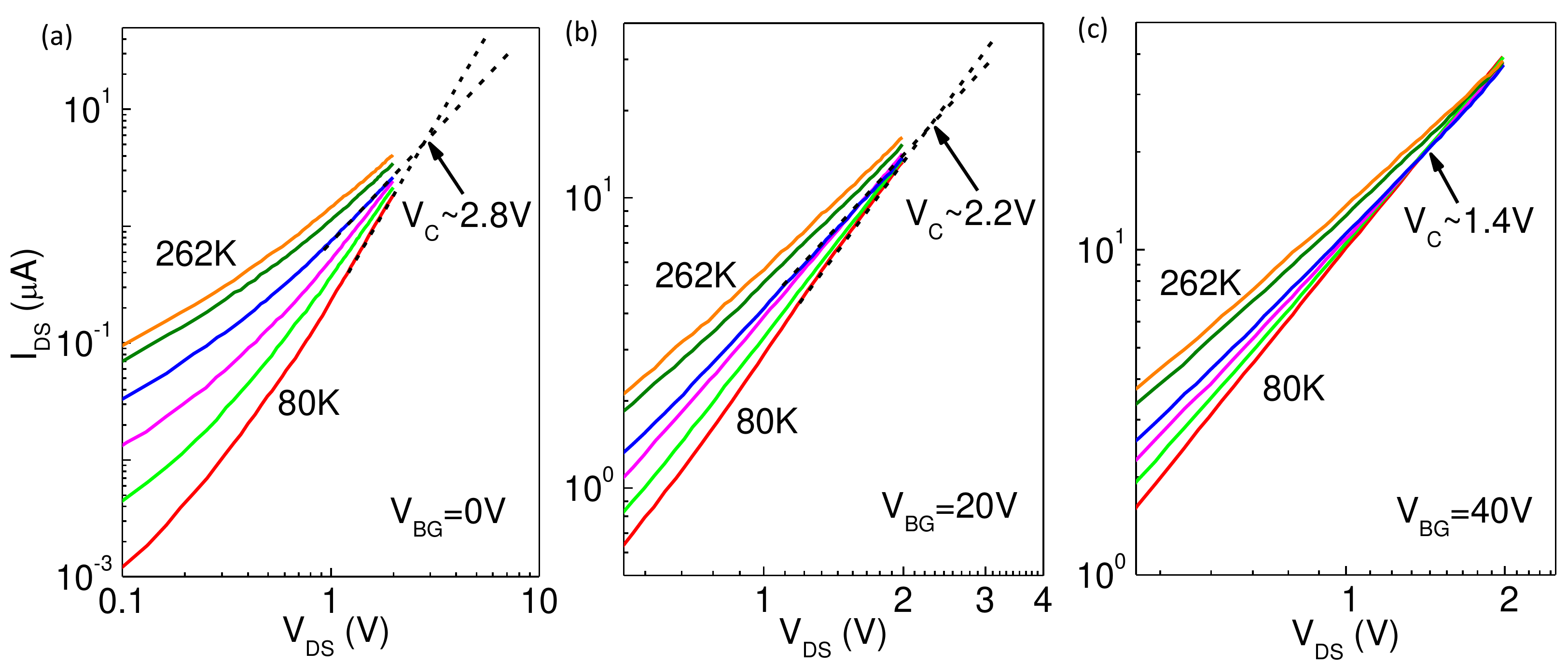}
\caption{Determination of $V_c$ from the temperature dependent $I-V$ characteristics for Dev~4 at (a)~$V_{BG}$~=~0~V, (b)~20~V, (c)~40~V. The
characteristics were taken at 80, 110, 140, 175, 230 and 262~K.}
\end{figure}
\end{center}

\newpage

%
\begin{center}
{\bf \LARGE Supplementary information}
\end{center}

\begin{center} {\bf \Large Trap-assisted space charge limited transport in short channel MoS$_2$ transistor}
\end{center}

\begin{center}
{Subhamoy Ghatak, and Arindam Ghosh}
\end{center}

\subsection{Transfer characteristics for all the devices:}

$I-V_{BG}$ characteristics of all the devices with $V_{DS}$~=~4~mV at $T~=~295~K$, showing large negative threshold voltage $V_T$ and high
$n$-doping.

 \begin{figure}[h]
 \centering
 \includegraphics[width=0.5\textwidth]{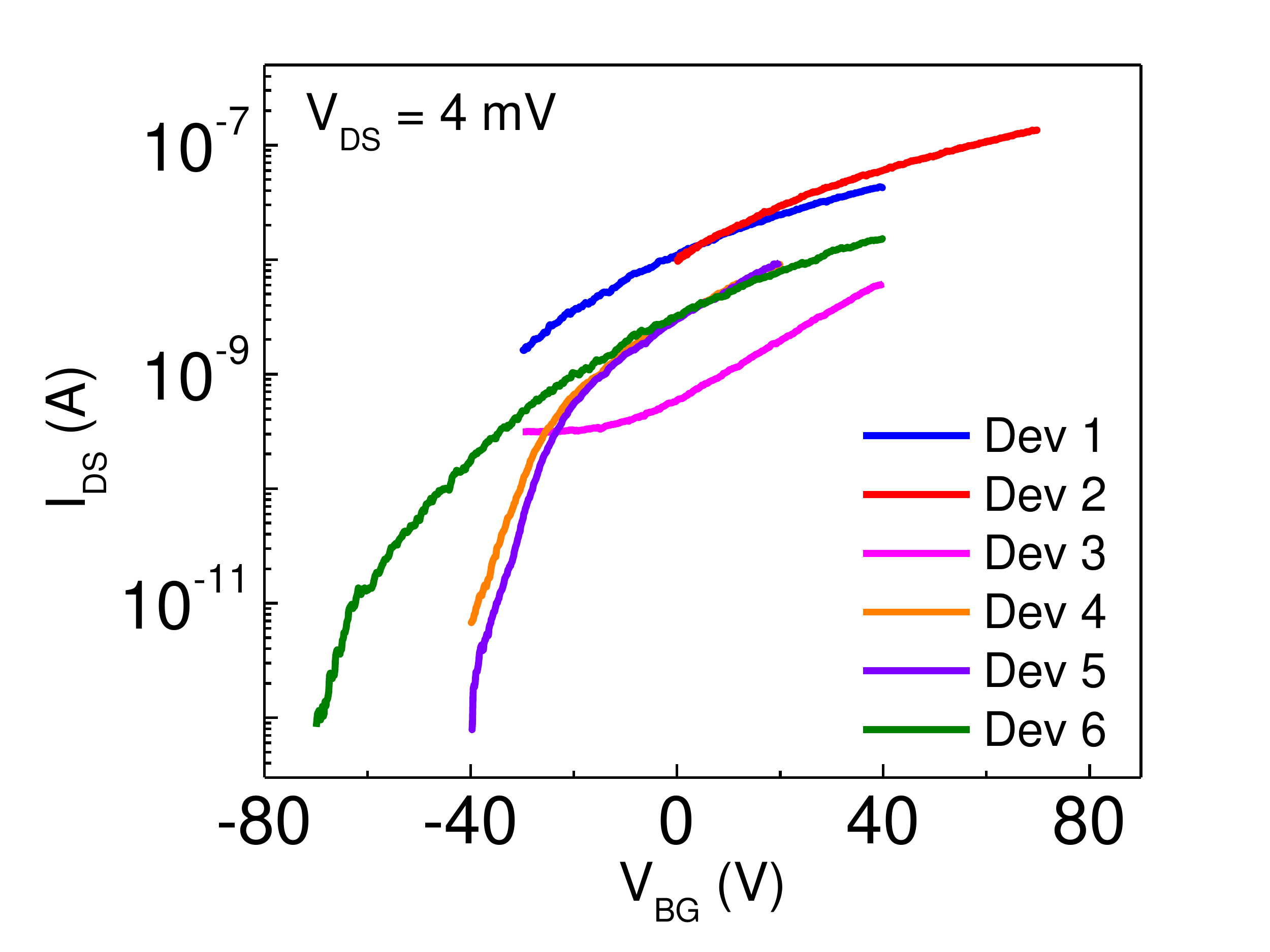}
 \caption{$I-V_{BG}$ characteristics of all the devices at 295~K for $V_{DS}$~=~4~mV.}
 \end{figure}

\subsection{Ohmic $I-V$ characteristics at low $V_{DS}$:}

 All the measurements were done with the devices which showed symmetric $I-V$ characteristics around $V_{DS}=0$ at all $T$ as a function of
 $V_{BG}$.

 \begin{figure}[h]
 \centering
 \includegraphics[width=0.5\textwidth]{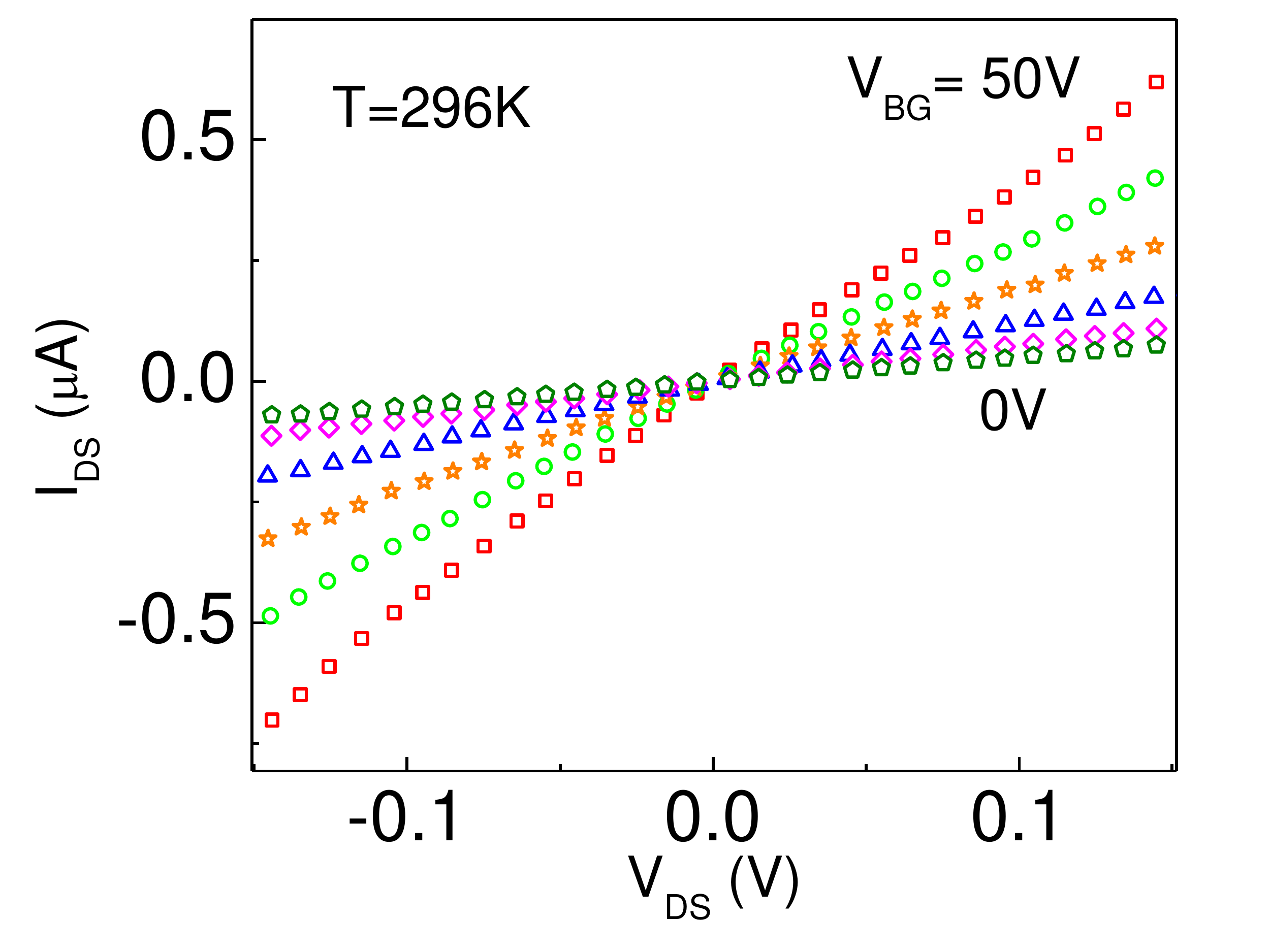}
 \caption{$I-V$ characteristics of a typical device near room temperature for $V_{DS\leq 100~mV}$ which show an ohmic conduction. }
 \end{figure}

\newpage
\subsection{Breakdown of channel at high $V_{DS}$:}

 We found the space charge limited conductivity regime after the ohmic regime with a sudden change in exponent of $V_{DS}$. With a further
 increase of $V_{DS}$, breakdown of the channel material was observed. For the device presented here the breakdown field was 2.7$\times10^5$
 V/cm.

 \begin{figure}[h]
 \centering
 \includegraphics[width=0.5\textwidth]{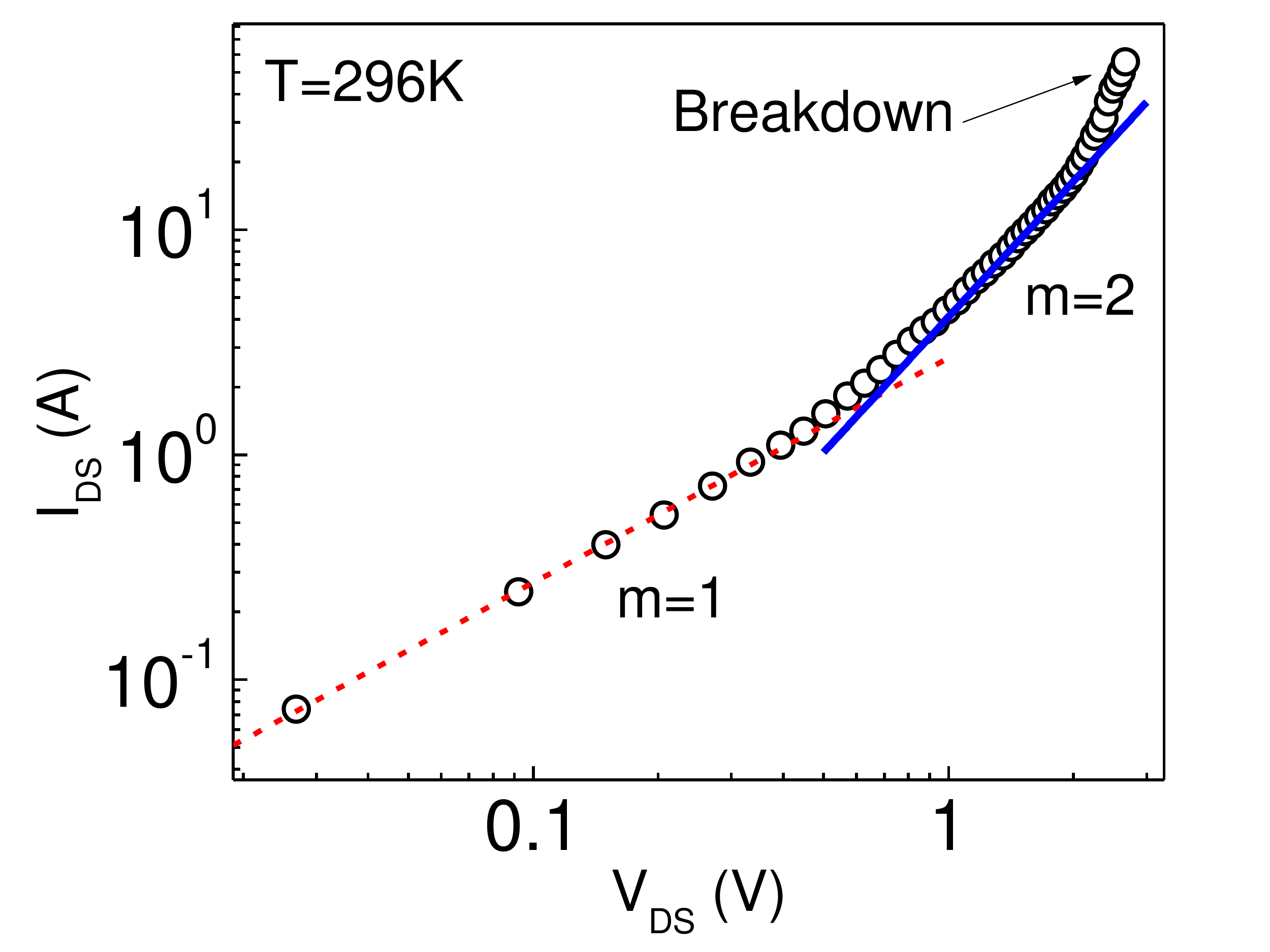}
 \caption{$I-V$ characteristics of a Dev 2 which showed breakdown at a high longitudinal electric field. }
 \end{figure}

\newpage
 \subsection{Representation of $I-V$ characteristics in semilog graph:}

The carrier transport mechanism for phonon-assistant tunneling, Schottky barrier manifest themselves with an exponential increase of current as
a function of $V_{DS}$. To eliminate effective contribution of those precesses in our device we plot the $I-V$ characteristics in a semilog
graph. Our data does not show a linear variation of current with $V_{DS}$.

 \begin{figure}[h]
 \centering
 \includegraphics[width=0.5\textwidth]{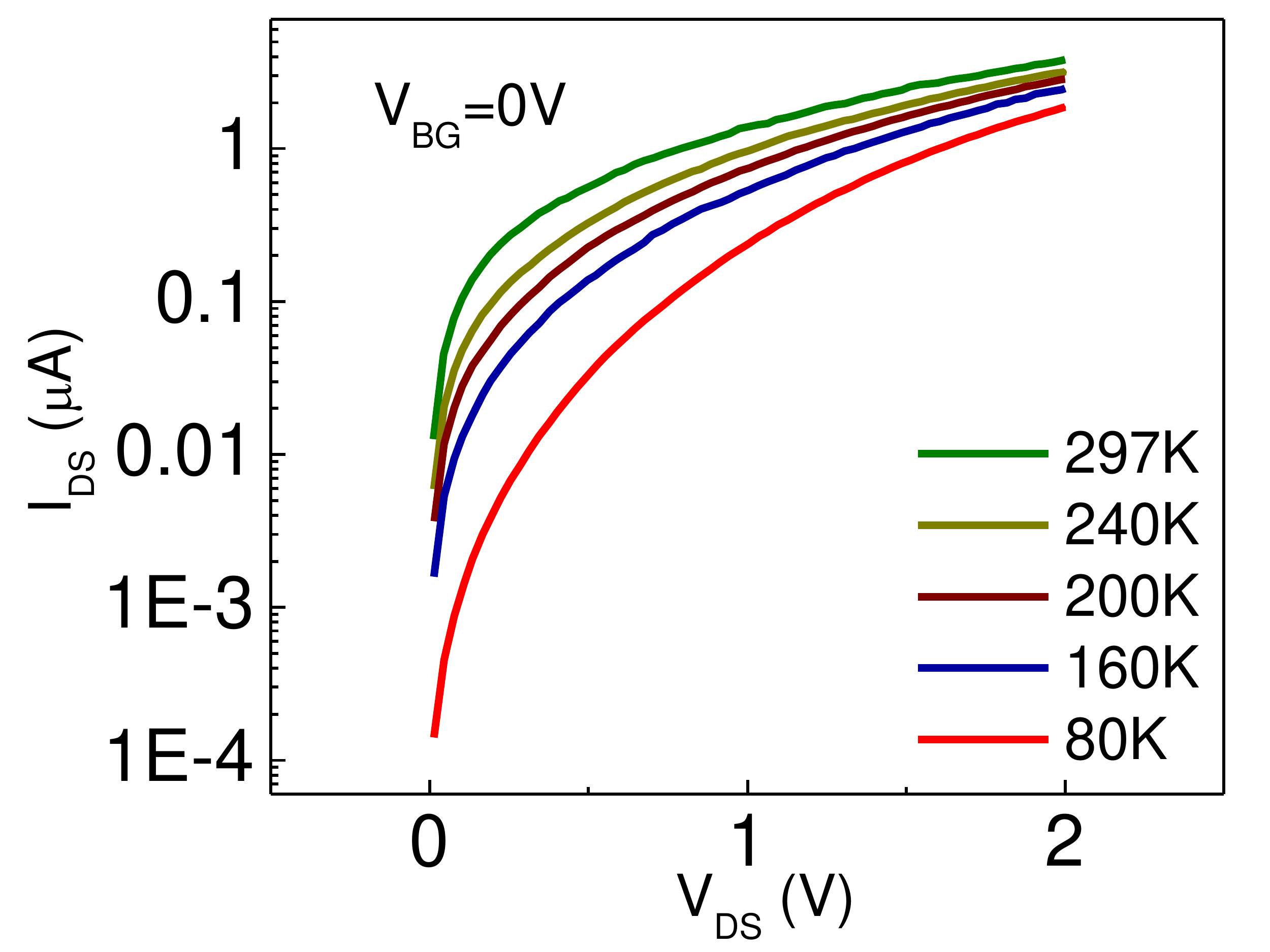}
 \caption{$I-V$ characteristics of a Dev 3 which showed breakdown at a high longitudinal electric field. }
 \end{figure}

\end{document}